\documentclass[floats,aps,amsfonts,notitlepage,superscriptaddress,twocolumn,eqsecnum,nofootinbib]{revtex4-1}

\usepackage[retainorgcmds]{IEEEtrantools}

\usepackage{ifpdf}
\usepackage[utf8]{inputenc}
\usepackage{graphicx}
\usepackage[colorlinks]{hyperref}
\usepackage{amsmath, amsthm, amssymb}
\usepackage{multirow}
\usepackage{url}
\usepackage{subfigure}
\usepackage{color}
\usepackage[usenames,dvipsnames,svgnames,table]{xcolor}
\usepackage{tabularx}

\allowdisplaybreaks[1]

\definecolor{CiteColor}{rgb}{0,0.5,0}
\hypersetup{citecolor=CiteColor}
\definecolor{RefColor}{rgb}{0.55,0,0}
\hypersetup{linkcolor=RefColor}
\definecolor{darkgreen}{rgb}{0.2,0.7,0.2}

\newcommand{\ab}{{\bar{a}}}
\newcommand{\bb}{{\bar{b}}}
\newcommand{\cb}{{\bar{c}}}

\newcommand{\hb}{{\bar{h}}}

\newcommand{\xb}{{\bar{x}}}
\newcommand{\tb}{{\bar{t}}}

\newcommand{\rb}{r_0}

\newcommand{\phb}{{\bar{\phi}}}
\newcommand{\thb}{{\bar{\theta}}}

\newcommand{\zrho}{{\rho}}

\newcommand{\adv}{{(adv)}}
\newcommand{\ret}{{(ret)}}
\newcommand{\sing}{{(S)}}
\newcommand{\reg}{{(R)}}

\DeclareMathOperator{\sgn}{sgn}

\newcommand{\mm}{\epsilon} 
\newcommand{\m}{\mu} 

\begin{document}
\title{Self-force regularization of a point particle for generic orbits in Kerr spacetime: electromagnetic and gravitational cases}

\author{Anna Heffernan}
\affiliation{IAC3--IEEC,  Universitat de les Illes Balears, E-07122 Palma, Spain}
\affiliation{Department of Physics, University of Florida, 2001 Museum Road, Gainesville, FL 32611-8440, USA}
\affiliation{School of Mathematics and Statistics, University College Dublin, Belfield, Dublin 4, Ireland}


\begin{abstract}
${}^*$Previous affiliations where part of this work was carried out.\\
The self-force is the leading method in modelling waveforms for extreme mass ratio inspirals, a key target of ESA's future space-based gravitational wave detector LISA. In modelling these systems, one approximates the smaller body as a point particle leading to problematic singularities that need to be removed. Modelling of this singular structure has settled on the Detweiler-Whiting singular field as the gold standard. As a solution to the governing wave equation itself, on removal, it leaves a smooth regular field that is a solution to the homogeneous wave equation, much like its well established flat spacetime counterpart. The mode-sum method enables subtraction of this singularity mode by mode via a spherical harmonic decomposition. The more modes one has, the faster the convergence in the $\ell$-sum, making these expressions highly beneficial, especially considering the heavy computational burden of waveform production. Until recently, only the two leading orders were known for generic orbits in Kerr spacetime. In a previous paper, we produced the next non-zero parameter for a scalar charged particle in curved spacetime, laying the groundwork for the electromagnetic and gravitational case which we present here.
\end{abstract}

\maketitle


\section{Introduction} \label{sec:intro}
Since their first detection \cite{Abbott:2016blz}, the network of ground-based gravitational detectors have gone from strength to strength. With approximately 90 detections to date \cite{LIGOScientific:2021djp}, they have ensured the future of gravitational wave astronomy. The ESA-led mission, LISA, is a future space-based gravitational wave detector, currently set to launch around 2034. This will open a new frequency window in gravitational waves, one where extreme mass ratio inspirals (EMRIs). EMRIs are binary compact body systems, usually treated as two black holes, a stellar mass black hole and a massive black hole of $\sim 10^7 - 10^9 M_{\odot})$, that live in the centre of galaxies. They are science rich systems, they can behave as dark standard sirens in constraining the Hubble constnat \cite{Laghi:2021pqk}, act as probes of massive black hole evolution via spin parameters \cite{Berti:2008af} and mass function constraints \cite{Gair:2010yu}, and with $\sim 10^4$ orbits of detectable signals, they will map out the Kerr spacetime, allowing precision tests of relativity \cite{Yagi:2009zm}.

The self-force solves for the motion of the system via a perturbation in the mass ratio in Einstein's field equations. The resulting wave equation is then solved via several numerical and semi-analytical methods, each of varying success. Arguably the most powerful is the matched expansions or Green function method, a semi-analytical method that involves solving directly for the regularised Green function in three overlapping regions via different techniques. This to-date has only been implemented in for a scalar particle with eccentric orbit in Schwarzschild \cite{Casals:2013mpa}. As is standard, near all self-force techniques are initially developed for a scalar particle in curved spacetime before progressing to an electric and eventually massive particle (as is the case with this work).

The effective source method where one uses the singular field to `effectively' source its wave equation, solving directly for the regular field, was independently developed by Vega and Detweiler \cite{Vega:2007mc}and Barack and Goldburn \cite{Barack:Golbourn:2007}. This has been successfully implemented for the gravitational self-force for circular orbits in both Schwarzschild \cite{Dolan:2012jg} and Kerr \cite{Isoyama:2014mja}. This leaves mode-sum, the technique we focus on here, first introduced by Barack and Ori \cite{Barack:1999wf} it uses a spherical decomposition to allow the safe subtraction of the singular field mode by mode. It has been fully implemented for generic orbits in both Scharzschild (eccentric) \cite{Barack:Sago:2010} and Kerr \cite{vandeMeent:2017bcc}. 

The mode-sum, when initially proposed, used the singular field that produced the MiSaTaQuWa equations \cite{Mino:Sasaki:Tanaka:1996, Quinn:1996am}. This singular field was later given an upgrade by Detweiler and Whiting \cite{Detweiler:2002mi} where they identified 'missing contributions' which when considered now gave a singular field that was a solution to the wave equation, much like its flat spacetime counterpart. Along with Messaritaki, they later implemented the mode-sum technique for this singular field \cite{Detweiler:2002gi} for circular obits of a scalar particle in Schwarzschild and with it several integration techniques which we adapt here. This paper also was the first to recognise that although, as Barack and Ori had observed, the sum of all parameters after the leading two sum to zero, as one cannot implement this sum over infinity, the higher parameters are still extremely useful in increasing the convergence of the $\ell$-sum. This observation led to a number of papers focused on systematically developing these higher orders, with Haas and Poisson extending the Detweiler et al.'s work to eccentric orbits \cite{Haas:2006ne}, Heffernan et al. extended this further to higher orders as well as the electromagnetic and gravitational cases for eccentric equatorial orbits in both Schwarzschild \cite{Heffernan:2012su} and Kerr \cite{Heffernan:2012vj}.

With recent mode-sum calculations now breaking into generic orbits in Kerr spacetime in scalar \cite{Nasipak:2021qfu} and gravity \cite{vandeMeent:2017bcc}, it seems appropriate to further extend these works to generic orbits in Kerr. The two leading parameters for generic orbits in Kerr were produced by Barack and Ori \cite{Barack:2002mh} in their initial pioneering work with details of their derivation arriving later in a self-force review by Barack \cite{Barack:2009}. The next non-zero parameter in generic Kerr was only recently produced by this author in \cite{Heffernan:2021olv}, hereon referred to as Paper I, for a scalar particle. Indeed, as that resulting expressions were 200 pages long (per component) in Mathematica, one can see why these expressions were not produced until now. Although large, for a scale of improved efficiency, that additional parameter resulted in code time reducing from a week to a day when used to investigate resonances \cite{Nasipak:2022xjh}. In addition \cite{Nasipak:2022xjh} was able to show even with numerical fitting, one cannot model resonances accurately enough without that additional parameter, thus making these parameters necessary for Kerr waveforms. To assist others in using those parameters, a Mathematica package was also released via Zenodo \cite{heffernan_anna_2022_6282572} and will be implemented in a future release of the black hole perturbation toolkit, bhptoolkit \cite{BlackHolePerturbationToolkit}. 

In producing the next non-zero parameter, we heavily use the techniques that were built in Paper I and hence refer to it often. In this paper, Sec.~\ref{sec:pert} gives the background on the governing equations for both an electric and massive particle in curved spacetime. It borrows heavily from Poisson et al.'s Living Review \cite{Poisson:2011nh} and should be seen as short summary of the necessary highlights. Sec.~\ref{sec:ssf} gives the techniques needed to produce expansions of the singular field; the base techniques are the same as Paper I with several necessary components identified that were not previously calculated. Sec.~\ref{sec:ms} gives a summary of the techniques and results of the mode-sum decomposition. These involve combining and adapting the coordinate rotation of Barack and Ori \cite{Barack:2002mha, Barack:2009} with the integration tricks of Detweiler et al. \cite{Detweiler:2002gi}. They are all described in detail in Paper I and so only a brief summary is given here. The resulting parameters are described here, as due to their lengthy nature, we provide them in a new release of the regularization Mathematica package on Zenodo \cite{heffernan_anna_2022_7071504} and shortly bhptoolkit \cite{BlackHolePerturbationToolkit}. Sec.~\ref{sec:dis} gives a summary and discussion of the paper's results.

\begin{figure} 
\includegraphics[width=6cm]{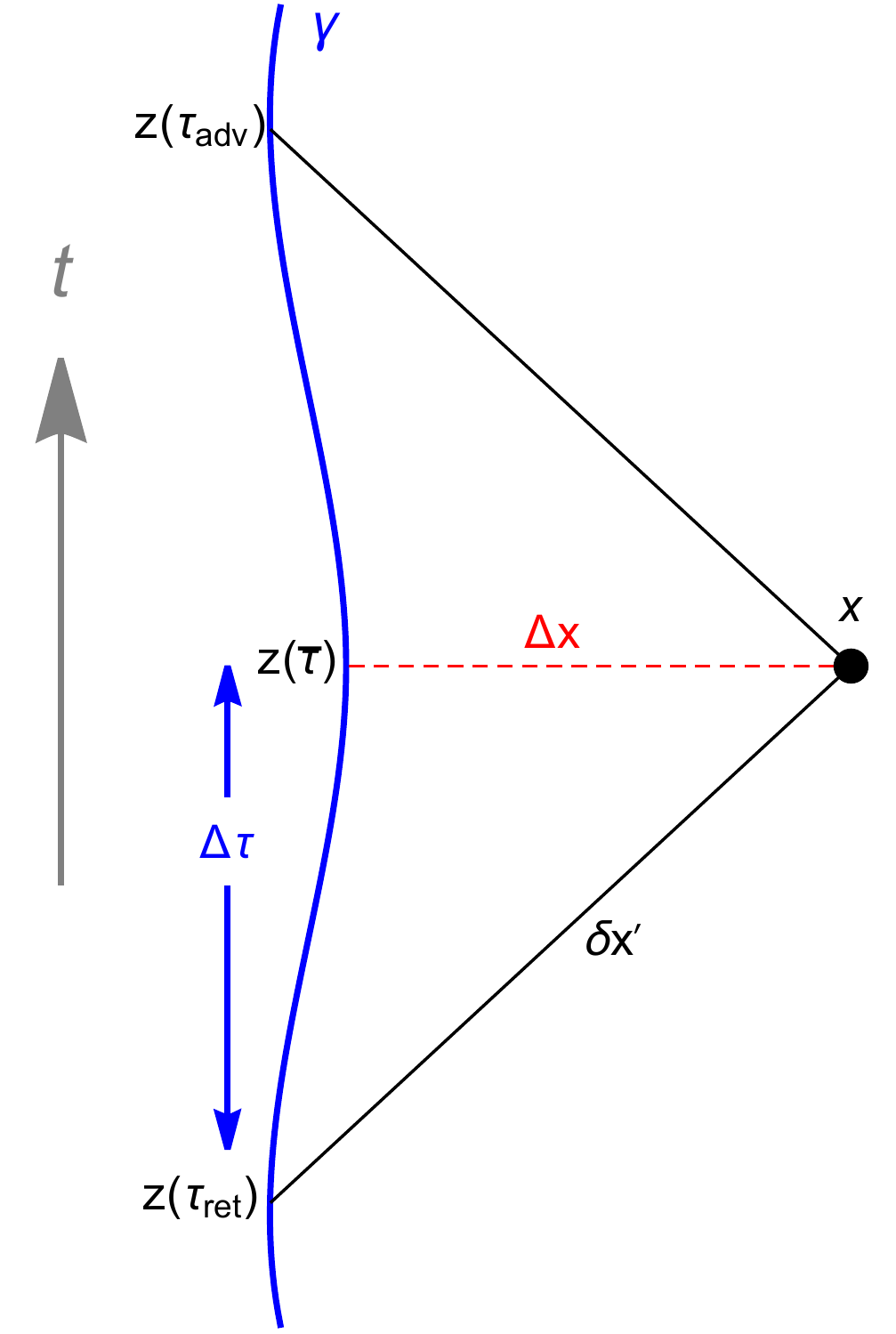} 
\caption{The worldline $\gamma$ of the point-particle source $z(\tau)$ where $x$ is the field point and $\xb \equiv z(\bar{\tau})$ is that point on the worldline that lives on the same constant time hypersurface as $x$ with $\Delta x=x-\xb$ as their spatial separation. $\delta x'=x-x'$ is the separation of the field point $x$ and a point $x'$ on the worldline (often taken to be the advanced $z^a(\tau_{\text{adv}})$ or retarded point $z^a(\tau_{\text{ret}})$ - those points on the worldline null-connected to the field point $x$).}
\label{fig:wl2D}
\end{figure}

Throughout this paper, we use units in which $G=c=1$ and adopt the sign conventions of
\cite{Misner:Thorne:Wheeler:1974}. We denote symmetrization of indices using parenthesis, $(ab)$, antisymmetrization using square brackets, $[ab]$, and exclude indices
from (anti)symmetrization by surrounding them by vertical
bars, $(a | b | c)$ and $[a | b | c]$. For spatial and four-velocity vectors we use the notation,
$x^a x^b \dots \equiv x^{a b c \dots}, u^a u^b \dots \equiv u^{ a b \dots}$ or $\dot{z}^{a b c \dots} \equiv \dot{z}^a \dot{z}^b \dot{z}^c \dots$, while biscalars with indices imply covariant differentiation, e.g., $\nabla^a \nabla^b \sigma(x,x') \equiv \sigma^{a b }$. We also make reference to several points; to clarify we direct the reader to Fig.~\ref{fig:wl2D} where:
\begin{itemize}
\item The worldline $\gamma$ of the source point particle is parametrized by proper time $\tau$ and described with position vector $x^{a'}\equiv z^a(\tau')$.
\item $x$ is field point off the worldline.
\item The point on $\gamma$ that lives on the same constant time spacelike hypersurface of $x$ is denoted $\xb \equiv z(\bar{\tau})$.
\item The retarded and advanced points on gamma that are null connected to $x$ are denoted $x^a_{\text{ret}} \equiv z^a(\tau_{\text{ret}}) \equiv z^a(\tau_{+}) $ and $x^a_{\text{adv}} \equiv z(\tau_{\text{adv}})\equiv z^a(\tau_{-}) $.
\item $\Delta x=x-\xb$ is the spacelike distance between the field point $x$ and $\xb$.
\item $\delta x'=x-x'$ is the distance between the field point $x$ and $x'$ on the worldine.
\item $\Delta \tau=\tau'-\bar{\tau}$ is the difference in proper time along the worldline between $\xb$ and another point $x'$, often chosen to be the retarded or advanced points.
\item We assume all distances are of similar length scale and introduce the dimensionless parameter $\epsilon \sim \Delta x \sim \delta x' \sim \Delta \tau$ to keep track of the order of each expansion.
\end{itemize}


\section{The Perturbative Self-Force } \label{sec:pert}


\subsection{An Electric Charged Particle Moving in Curved Spacetime} \label{sec:eomEM}

We consider an electric charged particle of charge $e$ moving in curved spacetime, on a worldline $\gamma$ parameterised by $\lambda$, with position vector $z^a(\lambda)$. In such a set-up, the charged particle will generate an electromagnetic field, $F_{a b}(x)=2A_{[a;b]}$ through its vector potential $A^a(x)$, with which it will then interact, hence the so-called self-force. The dynamics of such a system can be described by the action,
\begin{equation}
S=\int \left(\hat{\mathcal{L}}_{particle}+\hat{\mathcal{L}}_{field}+\hat{\mathcal{L}}_{interaction} \right) \sqrt{-g} d^4 x,
\end{equation}
where $\mathcal{L}$ are the Lagrangian densities
\begin{eqnarray} \label{eqn:Ls}
\hat{\mathcal{L}}_{particle}&=&-\mu \int_\gamma \delta_4 (x,z) d \tau, \\
\hat{\mathcal{L}}_{field}&=&-\frac{1}{16 \pi} F^{a b}(x) F_{a b}(x) ,\nonumber \\
\hat{\mathcal{L}}_{interaction}&=&e \int_\gamma A_a (x) g^a{}_b (x,z) \dot{z}^b \delta_4 (x,z) d \tau, \nonumber
\end{eqnarray}
$g^a{}_b (x,z)$ is the parallel propagator (more on this later), $\mu$ is the bare mass of the particle, g the determinant of the metric, $d \tau=\sqrt{-g_{ab} \dot{z}^a \dot{z}^b} d \lambda$ is the differential of proper time with overdot referring to differentiation with respect to $\lambda$, and $\delta_4(x,z)$ as defined in Eq.~(13.2) of \cite{Poisson:2011nh} is an invariant Dirac functional of curved spacetime,
\begin{equation}
\delta_4(x, x')=\frac{\delta_4(x-x')}{\sqrt{-g}} = \frac{\delta_4(x-x')}{\sqrt{-g'}}.  
\end{equation}

Varying this action with respect to the vector potential, $A^a$ and finding the stationary point yields the usual associated Euler-Lagrange equation,,
\begin{equation}
\frac{\partial \hat{\mathcal{L}}}{\partial A^a}-\nabla_a\left( \frac{\partial \hat{\mathcal{L}}}{\partial \dot{A}^a}\right)=0.
\end{equation}
Using the Lagrangian densities as defined in Eq.~\eqref{eqn:Ls} of the field and interaction terms results in Maxwell's equations,
\begin{eqnarray} \label{eqn:field}
F^{a b}{}_{;b}&=&4 \pi e \int_\gamma A_a (x) g^a{}_b (x,z) \dot{z}^b \delta_4(x,z(\tau)) d\tau, \\
	&\equiv& 4\pi j^a(x), \nonumber
\end{eqnarray}
where we have defined the current density,
\begin{equation} \label{eqn:j}
j^a (x) = e \int_\gamma g^a{}_b (x,z) \dot{z}^b  \delta_4(x,z(\tau)) d\tau.
\end{equation}
Using the fact that $F_{a b}$ is invariant under a gauge transformation, $A_a \rightarrow A_a + \Phi_{;a}$ for some arbitrary scalar function, $\Phi$, then $\Phi$ can always be chosen so that $A^a$ satisfies the Lorenze gauge,
\begin{equation}
A^a{}_{;a}=0.
\end{equation}
Thus Eq.~\eqref{eqn:field} emerges as the wave equation,
\begin{equation} \label{eqn:waveA}
\Box  A^a-R^a{}_b A^b=-4\pi j^a,
\end{equation}
where $R^a{}_b$ is the Ricci tensor.

For varying the action with respect to the position $z^a(\lambda)$, one need only consider the particle and interaction terms, resulting in a simplified system,
\begin{equation}
S=\int_\gamma \left(L_{particle}+L_{interaction} \right)d\lambda,
\end{equation}
where
\begin{eqnarray}
L_{particle}&=&-\mu \left[-g_{ab}(z) \dot{z}^a \dot{z}^b\right]^{1/2}, \\
L_{interaction}&=&e A_c (z) \dot{z}^c \left[-g_{ab}(z) \dot{z}^a \dot{z}^b\right]^{1/2}.
\end{eqnarray}
Again we arrive at the associated Euler-Lagrange equations,
\begin{equation}
\frac{\partial L}{\partial z^a}-\frac{d}{d \lambda}\left( \frac{\partial L}{\partial \dot{z}^a}\right)=0,
\end{equation}
which on taking $\lambda$ as proper time, lead to the equations of motion,
\begin{flalign} \label{eqn:eomEM}
F^a\equiv \mu \frac{D \dot{z}^a}{d \tau}  =e F^a{}_b (z)  \dot{z}^b.
\end{flalign}


\subsection{A Massive Particle Moving in Curved Spacetime} \label{sec:eomG}

The gravitational self-force approach tackles the Einstein field equations by exploiting the smallness of an object moving in a spacetime with a much larger length scale. For compact objects, this can translate to, $\m \sim \ell \ll M \sim \mathcal{L} \sim 1$ where $\m$ and $M$ are the massess of small and larger objects respectively, similarly $\ell$ and $\mathcal{L}$ represent their length scales.  It is, therefore, applicable to a `small' compact object (SCO), e.g., a stellar mass black hole, existing in a background spacetime described by a much larger object like a massive black hole (approximately $10^6$ solar masses), i.e., EMRIs.  If we consider the field generated by such a system, one can imagine far from the SCO, the field would be overwhelmed by the massive black hole, and the impact of the SCO's field can be seen as a perturbation, that is
\begin{equation} \label{eqn:g}
{\sf g}_{ab} = g_{ab} +  \mm h_{ab} +\mathcal{O}(\mm^2),
\end{equation}
where ${\sf g}$ describes the field of the system, $g$ describes the background metric associated with the massive black hole, $h$ the perturbation due to the SCO's field and $\mm$ represents the scaling $\m/\mathcal{L}$, that is the ratio of the SCO mass to the larger black hole length scale.  

Assuming the SCO represents the only mass in the system, the linearized Einstein field equations obtained using the metric Eq.~\eqref{eqn:g},
\begin{equation}
{\sf G}_{ab} [{\sf g}] = 8 \pi {\sf T}_{ab} [{\sf g}],
\end{equation}
now take the form,
\begin{equation}
G_{ab}[g] + \mm \delta G_{ab} [h] = \mm 8 \pi {\sf T}_{ab} [{\sf g}] + \mathcal{O}(\mm^2), 
\end{equation}
where ${\sf G}[{\sf g}]$ and $G [g]$ are the Einstein tensor defined with the metrics ${\sf g}$ and $g$ respectively, $\delta G [h]$ is the leading perturbation of the Einstein tensor while ${\sf T}[{\sf g}]$ 
represent the stress energy tensor of the SCO moving in the full 
spacetime, ${\sf g}$
.  By design, our background spacetime now has the freedom of being a solution to the Einstein vacuum field equations, i.e., $G_{ab}[g]=0$, allowing the Schwarzschild or Kerr solutions to describe our massive black hole as an uncharged non-spinning or spinning black hole respectively. 

To first order, the metric perturbation, $h$, is a solution to $\delta G_{ab}[h] = 8 \pi {\sf T}_{ab}[{\sf g}] + \mathcal{O} (\mm)$. This translates to
\begin{eqnarray} \label{eqn:gwave}
\Box \hb_{ab}+2 R_{acbd} \hb^{cd} - \hb_{b}{}^{c}{}_{;ca}- \hb_{a}{}^{c}{}_{;cb} + g_{ab} \hb^{cd}{}_{;cd} \nonumber \\ = -16 \pi {\sf T}_{ab}[{\sf g}] +\mathcal{O}(\mm),
\end{eqnarray}
where we have made use of the trace-reversed metric perturbation, $\hb^{ab}$, by means of the substitution,
\begin{equation}
h_{ab}=\hb_{ab}-\frac{1}{2} g_{ab} \hb^{c}{}_{c}.
\end{equation}
Employing the Lorenz gauge, $\hb^{ab}{}_{;b}=0$ Eq.~\eqref{eqn:gwave} now gives,
\begin{equation} \label{eqn:lin}
\Box \hb_{ab}+2 R_{acbd} \hb^{cd} = -16 \pi {\sf T}_{ab}[{\sf g}] + \mathcal{O} (\mm),
\end{equation}
where $\Box$ is the d'Alembertian operator with the background metric, $g$, and we have recovered the standard linearized Einstein field equations.  

As the SCO represents the only mass in the system, our stress energy tensor can be derived from the SCO's action functional,
\begin{equation}
S_p=-\m \int_\gamma \sqrt{-{\sf g}_{a'b'} \dot{z}^{a'} \dot{z}^{b'}} d\lambda'
\end{equation}
to give,
\begin{equation} \label{eqn:T}
{\sf T}^{ab} [{\sf g}] (x)= - \m \int_{\sf \gamma} \frac{{\sf g}^a{}_{a'} {\sf g}^b{}_{b'} {\sf \dot{z}}^{a'} {\sf \dot{z}}^{b'}}{\sqrt{- {\sf g}_{a'b'} {\sf \dot{z}}^{a'} {\sf \dot{z}}^{b'} }} \delta_4 (x, {\sf z}(\lambda')) d\lambda',
\end{equation}
where ${\sf z}(\lambda)$ describes the world line ${\sf \gamma}$ of the SCO in the full spacetime ${\sf g}$ (parameterized by $\lambda$) while ${\sf g}^a{}_{a'} \equiv {\sf g}^a{}_{a'} (x, {\sf z}(\lambda'))$ represents the bivector of parallel propagation (also in ${\sf g}$). Making use of the identities, 
\begin{eqnarray}
\lim_{x \rightarrow z'} g^a{}_{a';b} = 0,\quad
\left[g^b{}_{b'} \delta_4 (x,z(\lambda'))\right]_{;b} =\partial_{b'} \delta_4 (x,z(\lambda')), \nonumber
\end{eqnarray}
it can be shown,
\begin{eqnarray}
{\sf T}^{ab}{}_{;b}&[{\sf g}]& (x) = - \m \int_{\sf \gamma} \frac{{\sf g}^a{}_{a'} {\sf \dot{z}}^{a'} {\sf \dot{z}}^{b'}}{\sqrt{- {\sf g}_{a'b'} {\sf \dot{z}}^{a'} {\sf \dot{z}}^{b'} }} \partial_{b'}\delta_4 (x, {\sf z}(\lambda')) d\lambda', \nonumber\\
&=& \m \int_{\sf \gamma} \frac{d}{d \lambda'} \left[ \frac{{\sf g}^a{}_{a'} {\sf \dot{z}}^{a'}}{\sqrt{- {\sf g}_{a'b'} {\sf \dot{z}}^{a'} {\sf \dot{z}}^{b'} }} \right]\delta_4 (x, {\sf z}(\lambda')) d\lambda', \nonumber \\
&=& \m \int_{\sf \gamma} \frac{{\sf g}^a{}_{a'}}{\sqrt{- {\sf g}_{a'b'} {\sf \dot{z}}^{a'} {\sf \dot{z}}^{b'} }} \left[ \frac{{\sf D}  {\sf \dot{z}}^{a'} }{d \lambda'} - k'  {\sf \dot{z}}^{a'} \right]\delta_4 (x, {\sf z}(\lambda')) d\lambda', \nonumber
\end{eqnarray}
where ${\sf D}$ refers to covariant differentiaion with the full metric ${\sf g}$ and
\begin{equation} \label{eqn:k}
k = \frac{1}{\sqrt{- {\sf g}_{ab} {\sf \dot{z}}^{a} {\sf \dot{z}}^{b} }} \frac{d}{d \lambda} \left(\sqrt{- {\sf g}_{ab} {\sf \dot{z}}^{a} {\sf \dot{z}}^{b} }\right).
\end{equation}

If we enforce energy-momentum conservation, ${\sf T}^{ab}{}_{;b}=0$, we now have
\begin{equation} 
\frac{{\sf D}  {\sf \dot{z}}^{a} }{d \lambda} = k  {\sf \dot{z}}^{a}, \label{eqn:sfg}
\end{equation}
or
\begin{equation}
\frac{D  {\sf \dot{z}}^{a} }{d \lambda} =\left(\Gamma^{a}{}_{bc}-{\sf \Gamma}^{a}{}_{bc}\right) {\sf \dot{z}}^{b} {\sf \dot{z}}^{c}+k  {\sf \dot{z}}^{a},\label{eqn:sfa}
\end{equation}
where ${\sf \Gamma}$ and $\Gamma$ are the Christoffel symbols associated with covariant differentiation in ${\sf g}$ and $g$ respectively while $D$ refers to covariatiant differentiation with the background metric, $g$.  Perturbing Eq.~\eqref{eqn:sfa} via Eq.~\eqref{eqn:g}, and taking $\lambda$ to be the proper time of the background metric, now gives
\begin{eqnarray} 
a^{a [0]} &=& 0, \label{eqn:sfa0} \\
a^{a [1]} &=& -\frac{1}{2}\left(h^{a}{}_{b;c} + h^{a}{}_{c;b} - h_{bc}{}^{;a} - u^{ad} h_{bc;d} \right) u^{bc}, \nonumber \\
&=&-\frac{1}{2}\left(g^{ab}+u^{ab}\right) \left(2 h_{bc;d}-h_{cd;b}\right) u^{cd}, \label{eqn:sfa1}
\end{eqnarray}
where $a^{a[n]}$ is the $n^{\rm th}$ perturbation of the 4-acceleration and $k$, previously defined in Eq.~\eqref{eqn:k}, has become
\begin{eqnarray}
k &=& \frac{1}{\sqrt{1- h_{ab} u^{a b}}} \frac{d}{d \tau} \left(\sqrt{1-h_{ab} u^{a b}  }\right) ,\\
&=&\frac{1}{2} h_{ab;c} u^{abc} +\mathcal{O} (\mm^2).
\end{eqnarray}
Thus, by applying energy-momentum conservation, we have acquired a geodesic equation in the full spacetime, Eq.~\eqref{eqn:sfg}. From the prospective of the background spacetime, this is a geodesic at zero-th order by Eq.~\eqref{eqn:sfa0}, but at first order in Eq.~\eqref{eqn:sfa1}, it appears an acceleration or `force' has been applied; this is the so-called self-force.

We have glossed over the fact that we introduced a point particle assumption via Eq.~\eqref{eqn:T} in this derivation. The question whether this is a fair assumption has been thoroughly investigated, initially by \cite{Mino:Sasaki:Tanaka:1996} and \cite{Quinn:Wald:1997}, with elegant reviews in \cite{Poisson:2011nh} and \cite{Barack:2018yvs}.  In summary, the answer is yes, but only to first order. Second order requires a more structured source that can be thought of as a point particle equipped with the SCO's multipole moments \cite{Pound:2012dk}; this reduces to a point particle approximation at first order. We can therefore define our first order source term in Eq.~\eqref{eqn:gwave} as
\begin{equation} \label{eqn:Tg}
T_{ab} [g]= \m \int_\gamma g_{a' (a} g_{b) b'} u^{a'} u^{b'} \delta_4 (x, z(\tau')) d\tau',
\end{equation}
where we have explicitly illustrated $T$'s symmetry. With that our wave equation Eq.~\eqref{eqn:lin} becomes
\begin{equation} \label{eqn:waveh}
\Box \hb_{ab}+2 R_{acbd} \hb^{cd} = -16 \pi {\sf T}_{ab}[g] + \mathcal{O} (\mm).
\end{equation}


\subsection{Regularization} \label{sec:det}

As mentioned in Sec.~\ref{sec:intro}, if one wants to solve for the motion of a two-body system in full general relativity, one must deal with troublesome singularities. In the context of the self-force, we are concerned with the field singularity that emerges on the worldline of the smaller object, not the background spacetime, which is taken to be a vacuum.  We can see from both the electromagnetism and gravitational wave equations, Eqs.~\eqref{eqn:waveA} and \eqref{eqn:waveh} respectively, that these are sourced by delta functions.

As is standard with wave equations, the solutions of Eqs.~\eqref{eqn:waveA} and \eqref{eqn:waveh} are via Green functions. Our electromagnetic wave equation Eq.~\eqref{eqn:waveA}
\begin{equation}
\Box  A^a-R^a{}_b A^b=-4\pi j^a,
\end{equation}
clearly has the Green function solution,
\begin{eqnarray}  \label{eqn:fieldEM}
A^a(x)&=&\int_\gamma G^a{}_{b'} (x, x') j^{b'}(x') \sqrt{-g}d^4 x',
\nonumber \\
&=&e\int_\gamma G^a{}_{b'} (x, z(\tau')) u^{b'} d \tau',
\end{eqnarray}
where  we substituted Eq.~\eqref{eqn:j} and $G^a{}_{b'} (x, x')$ satisfies
\begin{equation} \label{eqn:emG}
\Box  G^a{}_{b'} (x, x')-R^a{}_b (x) G^b{}_{b'} (x, x')=-4\pi g^a{}_{b'}(x,x') \delta_4(x,x').
\end{equation}

Similarly for the gravitational case, the wave equation, Eq.~\eqref{eqn:waveh}
\begin{equation} 
\Box \hb_{ab}+2 R_{acbd} \hb^{cd} = -16 \pi {\sf T}_{ab}[g] + \mathcal{O} (\mm).
\end{equation}
has the Green function solution,
\begin{eqnarray} \label{eqn:fieldG}
\hb^{a b}{}_{;b'}(x)&=&\int_\gamma G^{a b}{}_{a' b'} (x, x') T^{a' b'}(x') \sqrt{-g}d^4 x',
\nonumber \\
&=&4  \mu \int_\gamma G^{a b}{}_{a' b'} (x, z(\tau')) u^{a' b'} d \tau',
\end{eqnarray}
where  we substituted Eq.~\eqref{eqn:Tg} and $G^{a b}{}_{a' b'} (x, x')$ satisfies
\begin{eqnarray} \label{eqn:gG}
\Box  G^{a b}{}_{a' b'} (x, x')+2R_c{}^a{}_d{}^b (x) G^{c d}{}_{a' b'} (x, x')
\nonumber \\
=-4\pi g^{(a}{}_{b'}(x,x') g^{b)}{}_{b'}(x,x')  \delta_4(x,x').
\end{eqnarray}

\begin{figure} 
\includegraphics[width=7cm]{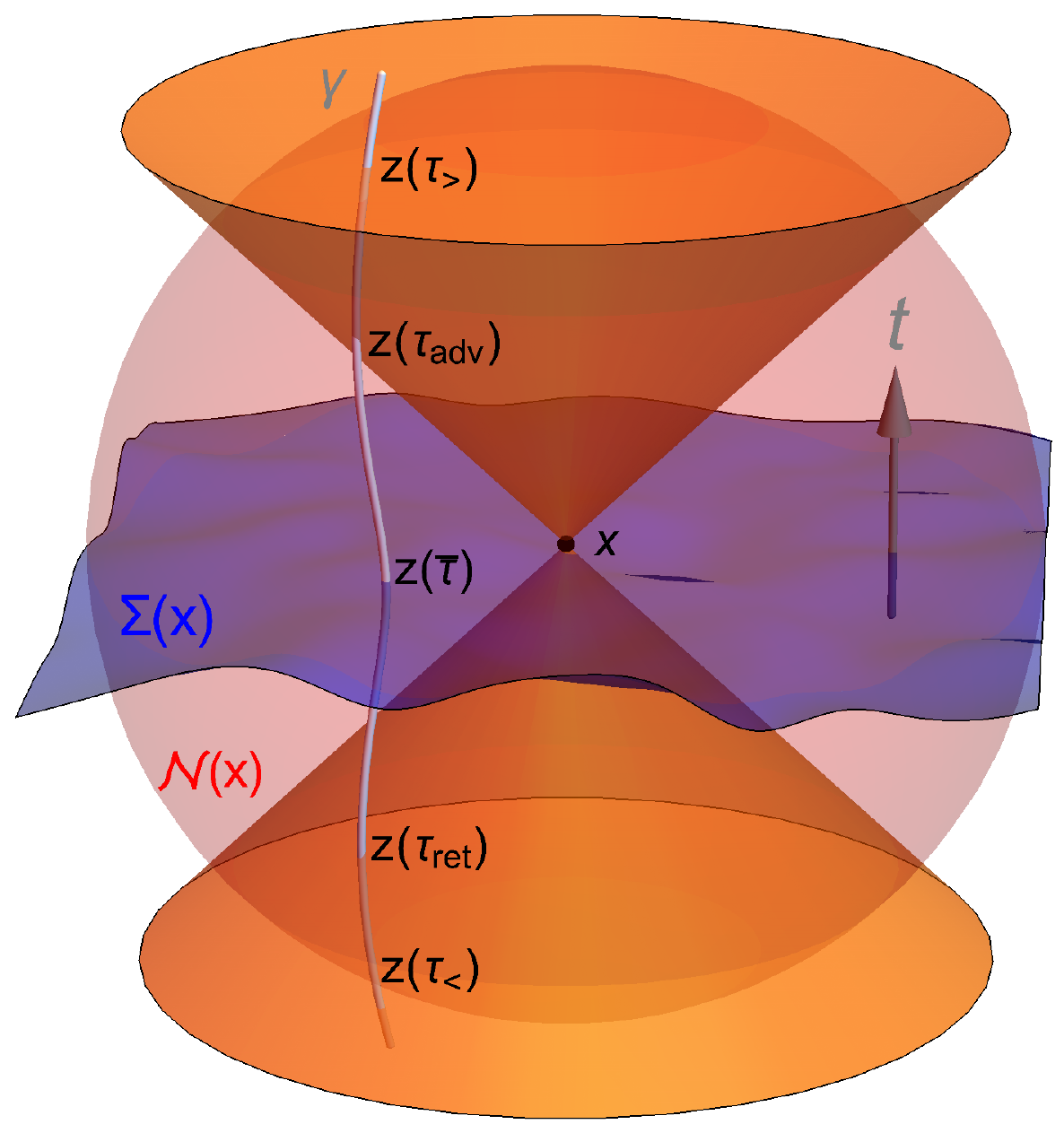} 
\caption{The worldline $\gamma$ of the source point particle $z(\tau')$ in curved spacetime passes through the local Normal neighbourhood of a field point $x$, $\mathcal{N}(x)$ so that its null connected retarded $z(\tau_{ret})$ and advanced $z(\tau_{adv})$ points also lie within $\mathcal{N}(x)$. The point $\xb = z(\bar{\tau})$ lives on the constant time spacelike hypersurface of the field point $\Sigma(x)$ with $\Delta x=x-\xb$ representing their separation.}
\label{fig:Gcurve}
\end{figure}

The well established method of tackling Green functions of the form Eqs.~\eqref{eqn:emG} and \eqref{eqn:gG} locally is to use the Hadamard ansatz \cite{Hadamard,DeWitt:1960},
\begin{equation} \label{eqn:greenEM}
G_{\pm} ^a{}_{b'} (x,x')=U^a{}_{b'} (x,x') \delta_{\pm} (\sigma) - V^a{}_{b'} (x,x') \Theta_{\pm} (-\sigma)
\end{equation}
for electromagnetism and for gravity,
\begin{equation} \label{eqn:greenG}
G_{\pm} ^{a b}{}_{a' b'} (x,x')=U^{a b}{}_{a' b'} (x,x') \delta_{\pm} (\sigma) - V^{a b}{}_{a' b'} (x,x') .\Theta_{\pm} (-\sigma)
\end{equation}
Here $\delta$ is the Dirac delta function, $\Theta$ the Heaviside step function, $\delta$ the Synge World function,`$+$' refers to the retarded solution and `$-$' the advanced while $U^a_{b'} (x,x')$ and $V^a_{b'} (x,x')$ are smooth bitensors. The Synge world function $\sigma$ in curved spacetime is
\begin{eqnarray}
\sigma (x'',x')&=&\frac12 \left( \lambda''-\lambda' \right) \int_{\lambda'}^{\lambda''} g_{ab}(z)  \frac{dz^a}{d\lambda}\frac{dz^b}{d\lambda} d\lambda,  \\
&=& \begin{cases}
		-\frac12 \Delta \tau^2 & \text{timelike } \beta,\\
		0  & \text{lightlike } \beta,\\
		\frac12 \Delta s^2 &\text{spacelike } \beta,
	\end{cases} 
\end{eqnarray}
The `$\pm$' attached to both $\delta$ and $\Theta$ in Eqs.~\eqref{eqn:greenEM} and \eqref{eqn:greenG} can be thought of as retarded and advanced versions of the Dirac delta function, $\Theta$ the Heaviside step function. To state this more mathematically, one considers a constant time spacelike hypersurface of the field point $\Sigma(x)$, then
\begin{equation}
\delta_\pm(\sigma(x,x'))=\Theta_\pm(\Sigma(x),x') \delta(\sigma(x,x')),
\end{equation} 
where $\Theta_\pm(\Sigma(x),x')$ is a step function, with  $\Theta_+(\Sigma(x),x')=1$ when $x'$ is in the past of $x$ and $\Theta_-(\Sigma(x),x')=1$ when it's in the future. Here it should be noted we are considering the local situation of fig.~\ref{fig:Gcurve}, where the field point $x$ and worldline $\gamma$ (or $x' \equiv z(\tau')$) are close enough so that both retarded and advanced points,  $z(\tau_{ret})$ and  $z(\tau_{adv})$, of $\gamma$ lie within $x$'s local neighbourhood, $\mathcal{N}(x)$. 

In flat spacetime, as was outlined in Paper I, one usually only has the direct part (null connected $U$'s) of the Green functions (a massive scalar field will have a tail part - $V$). To regularise, one forms a singular Green function that is the averaged sum of the retarded and advanced Green functions,
\begin{equation}
G^{(S) A}{}_{B'} = \frac12 (G^+{}^ A{}_{B'}+G^-{}^ A{}_{B'}),
\end{equation}
where we are using the notation that $A=a$ and $B'=b$ for electromagnetism while for gravity,  $A=a b$ and $B'=a' b'$. which is clearly a solution to Eqs.~\eqref{eqn:emG} and \eqref{eqn:gG} and hence has the identical singular structure as the both the retarded and advanced solutions. This is then subtracted from the retarded Green's function to give a smooth regular Green function that is a solution to the homogeneous wave equation.

However in curved spacetime, due to the tail term, such a singular field would depend on all the history and past of the source particle, therefore a different singular field is required. In Paper 1, we outlined the Quinn-Wald axiomatic approach that was used in the scalar case \cite{Quinn:2000wa}, and previously the electromagnetic and gravitational cases \cite{Quinn:1996am},  There, they used a number of axioms to validate subtracting the direct part of the flat spacetime singular field  $U^(A_{B')} (x,x') \delta (\sigma) $  and orbital averaging the remainder to regularise. This resulting field was regular and contained all the motion, it also was in agreement with derivation by Mino, Sasaki and Tanaka using asymptotic matching and solving in a buffer region \cite{Mino:Sasaki:Tanaka:1996}. The resulting equations of motion are since referred to as the MiSaTaQuWa equations.

These later got an upgrade via the Detweiler-Whiting singular field \cite{Detweiler:2002mi}, which showed that part of the tail term that was effectively being killed by orbital averaging could merely be subtracted along with the original singular field. Beautifully, as outlined in Paper I, adding this additional term to the singular Green function elevated it to a solution of the full wave equations (much like the flat spacetime scenario) and gave a much cleaner regularisation that did not require any orbital averaging.

\section{The Singular Self-Force} \label{sec:ssf}


\subsection{The Detweiler-Whiting singular field}
The Detweiler-Whiting singular Green function, 
\begin{equation} \label{eqn:GS}
G^\sing (x,x'){}^A{}_{B'} 
=\frac12 U{}^A{}_{B'}(x,x')\delta(\sigma)+ \frac12 V{}^A{}_{B'}(x,x')\Theta(\sigma),
\end{equation}
when integrated over the worldline via $\tau$ as depicted in Eqs.~\eqref{eqn:fieldEM} and \eqref{eqn:fieldG}, gives,
\begin{eqnarray}
A_a(x) &=&
 \frac{e}{2} \Bigg[ \frac{U_{a a'}(x,x')u^{a'}}{\sigma_{c'} u^{c'}} \Bigg]_{x'=x_{\rm \ret}}^{x'=x_{\rm \adv}}
    \\
  && + \frac{e}{2} \int_{\tau_{\rm \ret}}^{\tau_{\rm \adv}} V_{a a'} (x,z(\tau))u^{a'} d\tau,
  \nonumber \\
  \hb_{a b}(x) &=&
 2 \mu \Bigg[ \frac{U_{a b a' b'}(x,x')u^{a' b'}}{\sigma_{c'} u^{c'}} \Bigg]_{x'=x_{\rm \ret}}^{x'=x_{\rm \adv}}
    \\
  && + 2 \mu  \int_{\tau_{\rm \ret}}^{\tau_{\rm \adv}} V_{a b a' b'} (x,z(\tau))u^{a' b'} d\tau,
\nonumber
\end{eqnarray}
where the term $\sigma_{c'} u^{c'}$ simply arises from a change of variables in the integration $d \tau \rightarrow d\sigma$.

Expressions for the bitensors $U$ and $V$ can be obtained by substituting the singular Green function Eq.~\eqref{eqn:GS} into the governing equations, Eqs.~\eqref{eqn:emG} and \eqref{eqn:gG} and solving (reviews in \cite{Poisson:2011nh, Heffernan:2012xlf}). The direct $U$ bitensors are then given by,
\begin{eqnarray} \label{eqn:eqnUs}
U^{a b'} (x,x')&=&\Delta^{\tfrac12}(x,x') g^{a b'}(x,x'), \\
U^{a b a' b'} (x,x')&=&\Delta^{\tfrac12}(x,x') g^{(a |a'|}(x,x') g^{b)b'}(x,x'). \nonumber
\end{eqnarray}
where $\Delta^{\tfrac12}(x,x') $ is the Van Vleck-Morette determinant. The tail $V$ bitensors, when expressed as a formal expansion in the Synge world function, are shown to satisfy a recursion relationship initially sourced by the Van Vleck-Morette determinant  \cite{Decanini:Folacci:2005a}.

To obtain explicit expressions for these terms we carry out the same Riemann normal expansion as Paper I. We thus remind the reader the structure of this expansion, mainly
\begin{equation}
\Delta x = x - \xb \sim \epsilon, \qquad \delta x = x - x' \sim \epsilon,
\end{equation}
where $x$ is our field point, $\xb$ a point on the worldline on the same hypersurface as the field point $x$, $x'$ is a point anywhere on the world line where we use $x^\ab \equiv \xb^a$ and we have introduced $\epsilon$ as a small parameter to track orders. Using these coordinates in Paper I, we derived coordinate expansions for the Synge world function $\sigma$ and its derivatives $\sigma_a$, $\delta x$ the four-velocity on the worldline $u^{a'}$, along with the retarded and advanced points, $x^{a}(\tau_{ret})$ and $x^{a}(\tau_{adv})$ resectively, on the worldline. We also noted in \cite{Heffernan:2012su} it was shown that,
\begin{equation}
\Delta^{\tfrac12}(x,x') =1+\mathcal{O}(\epsilon)^4.
\end{equation}
We therefore have from Eq,~\eqref{eqn:eqnUs},
\begin{eqnarray}
U^{a b'} (x,x')&=&g^{a b'}(x,x') +\mathcal{O}(\epsilon)^4, \\
U^{a b a' b'} (x,x')&=&g^{(a |a'|}(x,x') g^{b)b'}(x,x')+\mathcal{O}(\epsilon)^4. \nonumber
\end{eqnarray}

One can see from these governing equations of both electromagnetism and gravity, most of the required expression were already determined via the Riemann Normal Expansion in Paper I. Indeed, the only missing ingredient for the direct part is the bivector of parallel transport, $g^a{}_{b'}$.  
As in Paper I, we Taylor expanded the metrics, i.e.,
\begin{eqnarray}
g_{a b} & \equiv & g_{a b} (x) 
= g_{\ab \bb} +  g_{\ab \bb, \cb} \Delta x^{\cb} \epsilon + \mathcal {O} (\epsilon)^2,  \\
g_{a' b'} &\equiv& g_{a b} (x') 
= g_{\ab \bb} +  g_{\ab \bb, \cb} \left(\Delta x - \delta x \right)^{c} \epsilon + \mathcal {O} (\epsilon)^2. \nonumber
\end{eqnarray}
Then similar to solving for $\sigma$ in Paper I, we define our $g^a{}_{b'} (x,x')$ as an expansion with unknown coefficients, labelled $G$,
\begin{eqnarray}
g^a{}_{b'} &=& 
\delta^a{}_{b} + \epsilon G^a{}_{b c} \delta x^{c'} + \epsilon^2 G^a{}_{b c d} \delta x^{c'} \delta x^{d'}  \\
&&
+ \epsilon^3 G^a{}_{b c d e} \delta x^{c'} \delta x^{d'} \delta x^{e'} + \mathcal{O} (\epsilon)^4, \nonumber 
\end{eqnarray} 
and further expand with  $x \rightarrow \xb + \Delta x$. Using $g^a{}_{b';c'} \sigma^{c'}=0$ as our governing equation for the bivector of parallel transport, we solve for the $G$'s, 
\begin{eqnarray}
G^a{}_{b c}&=&-\Gamma^\ab{}_{\bb \cb}, \\
G^b{}_{c e f}&=&\tfrac{1}{4}  (\Gamma ^{b}{}_{e}{}^{i} \Gamma_{icf} 
+ \Gamma ^{b}{}_{c}{}^{i} \Gamma _{ief} + \Gamma _{ief} \Gamma ^{ib}{}_{c} 
\nonumber \\
&&
-  \Gamma _{icf} \Gamma ^{ib}{}_{e} -  \Gamma _{cef}{}^{,b} 
-  \Gamma_{ecf}{}^{,b} + \Gamma ^{b}{}_{ef}{}_{,c} 
\nonumber \\
&&
+ g^{bi} \Gamma_{efi}{}_{,c} + \Gamma^{b}{}_{ce}{}_{,f} 
+ g^{bi} \Gamma_{cei}{}_{,f})
\big)
\end{eqnarray}
\begin{widetext}
\begin{eqnarray}
G^b{}_{c i j k}&=&
\frac1{24} \bigl\{
2 \Gamma {ci}{}^{n} \Gamma_{nko} \Gamma ^{ob}{}_{j} 
+ 2 \Gamma _{ci}{}^{n} \Gamma _{okn} \Gamma ^{ob}{}_{j} 
+ 2 \Gamma _{jk}{}^{n} \bigl[
	\Gamma _{nco} \Gamma ^{ob}{}_{i} 
	+ \Gamma _{ocn} \Gamma ^{ob}{}_{i} 
	-  (\Gamma ^{b}{}_{no} + \Gamma _{n}{}^{b}{}_{o}) \Gamma ^{o}{}_{ci}
\bigr]
- 4 \Gamma^{nb}{}_{i} \Gamma _{okn} \Gamma{o}{}_{cj} 
\nonumber \\
&& \quad
+ \Gamma ^{b}{}_{c}{}^{n} \Gamma_{okn} \Gamma ^{o}{}_{ij} 
+ \Gamma ^{nb}{}_{c} \Gamma _{okn} \Gamma ^{o}{}_{ij} 
+ \Gamma ^{b}{}_{no} \Gamma _{ci}{}^{n} \Gamma ^{o}{}_{jk} 
+ \Gamma ^{b}{}_{no} \Gamma _{ic}{}^{n} \Gamma ^{o}{}_{jk} 
-  \Gamma ^{b}{}_{c}{}^{n} \Gamma _{ino} \Gamma ^{o}{}_{jk} 
+ \Gamma _{ci}{}^{n} \Gamma _{n}{}^{b}{}_{o} \Gamma ^{o}{}_{jk} 
\nonumber \\
&&  \quad
+ \Gamma _{ic}{}^{n} \Gamma _{n}{}^{b}{}_{o} \Gamma ^{o}{}_{jk} 
-  \Gamma _{ino} \Gamma ^{nb}{}_{c} \Gamma ^{o}{}_{jk} 
- 2 \Gamma ^{b}{}_{no} \Gamma ^{n}{}_{ci} \Gamma ^{o}{}_{jk} 
+ 2 \Gamma _{n}{}^{b}{}_{o} \Gamma^{n}{}_{ci} \Gamma ^{o}{}_{jk} 
- 2 \Gamma ^{n}{}_{ci} \Gamma _{o}{}^{b}{}_{n} \Gamma ^{o}{}_{jk} 
+ 2 \Gamma ^{nb}{}_{i} \Gamma _{ocn} \Gamma ^{o}{}_{jk} 
\nonumber \\
&&  \quad
-  \Gamma ^{n}{}_{ij} \Gamma _{ckn}{}^{,b} 
+ 2 \Gamma ^{n}{}_{ci} \Gamma _{jkn}{}^{,b} 
-  \Gamma ^{n}{}_{ij} \Gamma _{nck}{}^{,b} 
+ 2 \Gamma ^{n}{}_{ci} \Gamma_{njk}{}^{,b} 
+ 2 \Gamma _{cij}{}_{,k}{}^{,b} 
+ 2 \Gamma _{icj}{}_{,k}{}^{,b} 
- 2 \Gamma _{ij}{}^{n} \Gamma ^{b}{}_{kn}{}_{,c} 
- 2 \Gamma ^{n}{}_{ij} \Gamma ^{b}{}_{kn}{}_{,c} 
\nonumber \\
&& \quad
+ 2 \Gamma ^{nb}{}_{i} \Gamma _{jkn}{}_{,c}
 -  \Gamma ^{n}{}_{ij} \Gamma _{k}{}^{b}{}_{n}{}_{,c}
  - 2 \Gamma _{ij}{}^{n} \Gamma _{n}{}^{b}{}_{k}{}_{,c}
  -  \Gamma ^{n}{}_{ij} \Gamma_{n}{}^{b}{}_{k}{}_{,c} 
  + 2 \Gamma^{nb}{}_{i} \Gamma_{njk}{}_{,c} 
  - 2 \Gamma _{ci}{}^{n} \Gamma^{b}{}_{jn}{}_{,k}
   - 2 \Gamma ^{b}{}_{c}{}^{n} \Gamma _{ijn}{}_{,k} 
   \nonumber \\
   &&  \quad
   - 2 \Gamma ^{nb}{}_{c} \Gamma _{ijn}{}_{,k} 
   - 2 \Gamma_{ci}{}^{n} \Gamma _{n}{}^{b}{}_{j}{}_{,k} 
   + 2 \Gamma ^{b}{}_{i}{}^{n} \bigl[
   	\Gamma _{jk}{}^{o} (
		\Gamma _{nco} + \Gamma _{ocn}
	) 
	+ \Gamma _{cj}{}^{o} (
		\Gamma _{nko} + \Gamma _{okn}
	) 
	+ 2 \Gamma _{nko} \Gamma ^{o}{}_{cj} 		
	+ 2 \Gamma _{okn} \Gamma ^{o}{}_{cj} 
\nonumber \\
&& \qquad
	+ \Gamma _{nco} \Gamma ^{o}{}_{jk} 
	+ \Gamma _{ocn} \Gamma ^{o}{}_{jk} 
	- \Gamma _{jkn}{}_{,c} -  \Gamma _{njk}{}_{,c} 
	- 2 \Gamma _{cjn}{}_{,k} 
	- 2 \Gamma_{ncj}{}_{,k}
\bigr]
- 2 \Gamma ^{b}{}_{c}{}^{n} \Gamma _{nij}{}_{,k} 
- 2 \Gamma ^{nb}{}_{c} \Gamma _{nij}{}_{,k} 
- 2 \Gamma ^{b}{}_{ij}{}_{,c}{}_{,k} 
\nonumber \\
&&  \quad
- 2 \Gamma _{i}{}^{b}{}_{j}{}_{,c}{}_{,k} 
- 2 \Gamma ^{b}{}_{ci}{}_{,j}{}_{,k} 
- 2 \Gamma _{c}{}^{b}{}_{i}{}_{,j}{}_{,k} 
- 2 \Gamma ^{n}{}_{ci} \Gamma ^{b}{}_{jk}{}_{,n} 
+ \Gamma ^{n}{}_{ij} \Gamma _{c}{}^{b}{}_{k}{}_{,n} 
- 4 \Gamma ^{nb}{}_{i} \Gamma _{cjk}{}_{,n} 
- 2 \Gamma ^{n}{}_{ci} \Gamma _{j}{}^{b}{}_{k}{}_{,n} 
- 4 \Gamma^{nb}{}_{i} \Gamma _{jck}{}_{,n} 
\nonumber \\
&&  \quad
+ \Gamma ^{n}{}_{ij} \Gamma _{k}{}^{b}{}_{c}{}_{,n}
\bigr\}
\end{eqnarray}
\end{widetext}

As mentioned previously, to obtain an expression for the tail bitensor $V$, it is written as an expansion in $\sigma$, 
\begin{eqnarray}
V^A{}_{B'}(x,x')&=& \sum_{n=0}^{\infty} V_n^A{}_{B'}(x,x') \sigma^n(x,x'), \\
&=&  V_0^A{}_{B'}(x,x') + \mathcal{O}(\epsilon)^2,
\end{eqnarray}
with the $V_n$'s solving a recursion relations and $ \sigma^n$ begin $\sigma$ to the power of $n$ and not a contravariant derivative. The second line simply follows from $\sigma$ having leading order $\epsilon^2$. In Paper I we noted, in the scalar case $V_0=\mathcal{O}(\epsilon)^2$, and so it was unnecessary to calculate any $V$ terms. Similarly as shown in the Appendix of \cite{Heffernan:2012su}, this is also the scenario for the electromagnetic case. However, for gravity, it is not so we also need to calculate the $V$ terms for gravity.

We only need $V_0$ and thus the initial equation sourced by the Van Vleck-Morette determinant will suffice, 
\begin{eqnarray}
2 \sigma^{;c'} V^{aba'b'}_{0}{}_{;c'} - 2 V^{aba'b'}_0 \Delta^{-\tfrac{1}{2}} \sigma^{;c'} \Delta^{\tfrac{1}{2}}{}_{;c'} 
+ 2 V^{aba'b'}_0  \\
+ \left(\delta^{a'}{}_{c'}\delta^{b'}{}_{d'}\square' + 2 R^{a'}{}_{c'}{}^{b'}{}_{d'} \right) \left(\Delta^{\tfrac{1}{2}} g^{c'(a}g^{b) d'}\right) = 0. \nonumber
\end{eqnarray}
We then use the expansion 
\begin{equation} \label{eqn: v0 G}
V^{aba'b'}_0 (x, x') = v^{aba'b'}_{0}(x) + v^{aba'b'}_{0}{}_{c} (x) \delta x^{c'} +\mathcal{O}(\epsilon)^2
\end{equation}
where  $v^{aba'b'}_{0}$ and $ v^{aba'b'}_{0}{}_{c} $are unknown coefficients for which we solve,
\begin{widetext}
\begin{eqnarray}
v^{bcd'e'}_{0}(x) &=&
\tfrac{1}{4} \Bigl\{
	-2 \Gamma ^{def} \Gamma _{f}{}^{bc} 
	+ \Gamma ^{dcf} \Gamma _{f}{}^{be} 
	+ 2 \Gamma _{f}{}^{de} \Gamma ^{fbc} 
	+ \Gamma _{f}{}^{ce} (\Gamma ^{dbf} -  \Gamma ^{fbd}) 
	-  \Gamma _{f}{}^{cd} \Gamma ^{fbe} 
\nonumber \\
&& \quad
	+ g^{bf} g^{ci} \bigl[
		-2 \Gamma ^{d}{}_{fi}{}^{,e} 
		+ g4^{ej} \bigl(
			\Gamma _{fij}{}^{,d} + \Gamma _{ifj}{}^{,d} 
			- 2 \Gamma _{jfi}{}^{,d} + \Gamma ^{d}{}_{ij}{}_{,f} 
			+ \Gamma ^{d}{}_{fj}{}_{,i}
		\bigr) 
\nonumber \\
&& \qquad
		+ g^{dj} g^{ek} \bigl(
			\Gamma _{kij}{}_{,f} + \Gamma _{kfj}{}_{,i} 
			-  \Gamma _{fij}{}_{,k} -  \Gamma _{ifj}{}_{,k}
		\bigr)
	\bigr]
\Bigr\},
\end{eqnarray}
\end{widetext}
where we don't include $v^{aba'b'}_{0}{}_{c}$ due to its size.

Putting this together with the expressions from Paper I gives us coordinate expansions for $A^a(x)$ and $\hb^{a b} (x)$. To obtain the self-force, we use the equations of motion Eqs.~\eqref{eqn:sfa1} and \eqref{eqn:eomEM} outlined in Sec.~\ref{sec:pert}, that is
\begin{eqnarray}
F^a_{(\text{EM}) }=e F^a{}_b (z)  \dot{z}^b, \\
F^a_{(\text{G})} =\mu \, k^{abcd} \hb^{\rm{\reg}}_{bc; d},
\end{eqnarray}
where we have rearranged the gravitational case as is standard by introducing,
\begin{eqnarray}
k^{abcd} &\equiv& \frac12 g^{ad} u^b u^c - g^{ab} u^c u^d - 
  \frac12 u^a u^b u^c u^d \nonumber \\
  &&+ \frac14 u^a g^{b c} u^d + 
 \frac14 g^{a d} g^{b c}.
\end{eqnarray}
Similar to Paper I, we find the self-force has the structure,
\begin{widetext}
\begin{equation} \label{eqn:fasum}
F^{\rm{\sing}}_a \left(x\right) = \sum_{n=-1}^{\infty} \sum_{p=-n-2}^{\lfloor (n-1)/2 \rfloor}b^{[n]}_{\ab \bar{c}_1 \dots \bar{c}_{(n-2p)}}(\bar{x}) \Delta x^{c_1} \dots \Delta x^{c_{(n-2p)}} \zrho^{2 p - 1} \epsilon^{n-1},
\end{equation}
\end{widetext}
where
\begin{equation} \label{eqn:rhoS}
\rho^2=(g_{\ab \bb}+u_{\ab \bb}) \Delta x^{ab}
\end{equation}
and with the caveat that for $n=2$ the limit on the $p$ sum becomes $\lfloor n/2 \rfloor$. To see this explicitly, we include the first two orders in the electromagnetic case,
\begin{widetext}
\begin{eqnarray}
F^{(S)}_{a[-1]}&=&\tfrac{1}{2} \Delta x_{b} + \tfrac{1}{2} u_{b} u^{c} \Delta x_{c},
 \\
F^{(S)}_{a[0]}&=&
 \frac1{4 \rho^3} \Bigl\{
 	\Gamma _{bcd} \Delta x^{c d} 
	+ 2 \Gamma _{cbd} \Delta x^{c d}
	+ u^{c} \bigl[
		\Gamma _{cde} (2 u^{d} \Delta x_{b} + u_{b} \Delta x^{d}) 
		+ 2 (\Gamma _{dbe} u^{d} \Delta x_{c} + \Gamma _{dce} u_{b} \Delta x^{d})
\bigr] \Delta x^{e}
\Bigr\}
\nonumber \\
&&- \frac{3}{4 \rho^5} \Bigl[
	\Gamma _{cde} \Delta x_{b } \Delta x^{c d e} 
	+ u^{c} \Delta x_{c} \Delta x^{f} \bigl(
		\Gamma _{def} (u^{d} \Delta x_{b} 
		+ u_{b} \Delta x^{d}) \Delta x^{e} 
		+ \Gamma _{efi} u_{b} u^{d e}  \Delta x_{d i} 
	\bigr)
\Bigr].
\end{eqnarray}
\end{widetext}


\section{Mode-sum Decomposition} \label{sec:ms}
\subsection{Coordinate rotation and integration techniques}
Once owe have an expression for the singular self-force, we follow the mode-sum method of Barack and Ori  \cite{Barack:1999wf} and regularise by decomposing into spherical harmonics and subtracting mode by mode,
\begin{equation} \label{eqn:Fa}
F_a(\xb)=\sum^\infty_{\ell}\left[F_a^{\ell \rm{\ret}}(\xb) - F_a^{\ell \rm{\sing}} (\xb)\right],
\end{equation}
where
\begin{eqnarray} \label{eqn:FRS}
F_a^{\ell\rm{\sing}}(\xb) &=& \sum_{m} F_a^{\ell m \rm{\sing}} (\tb, \rb) Y_{\ell m} (\bar{\theta}, \bar{\phi}), \nonumber \\
&=& \sqrt{\frac{2\ell+1}{4\pi}} F_a^{\ell 0\rm{\sing}} (\tb, \rb), \nonumber \\
&=&\lim_{\Delta r \rightarrow 0} \frac{2\ell+1}{4 \pi}  \int F_a^{\rm{\sing}} (0,\Delta r,\alpha,\beta) \nonumber \\
&& \quad \times P_\ell(\cos{\alpha}) d \Omega.
\end{eqnarray}
To carry out this integration, we follow the same technique as outlined in detail in Paper I. For completeness we give a brief overview here but refer the reader to Paper I for full details. 

\begin{figure}
\includegraphics[width=6.5cm]{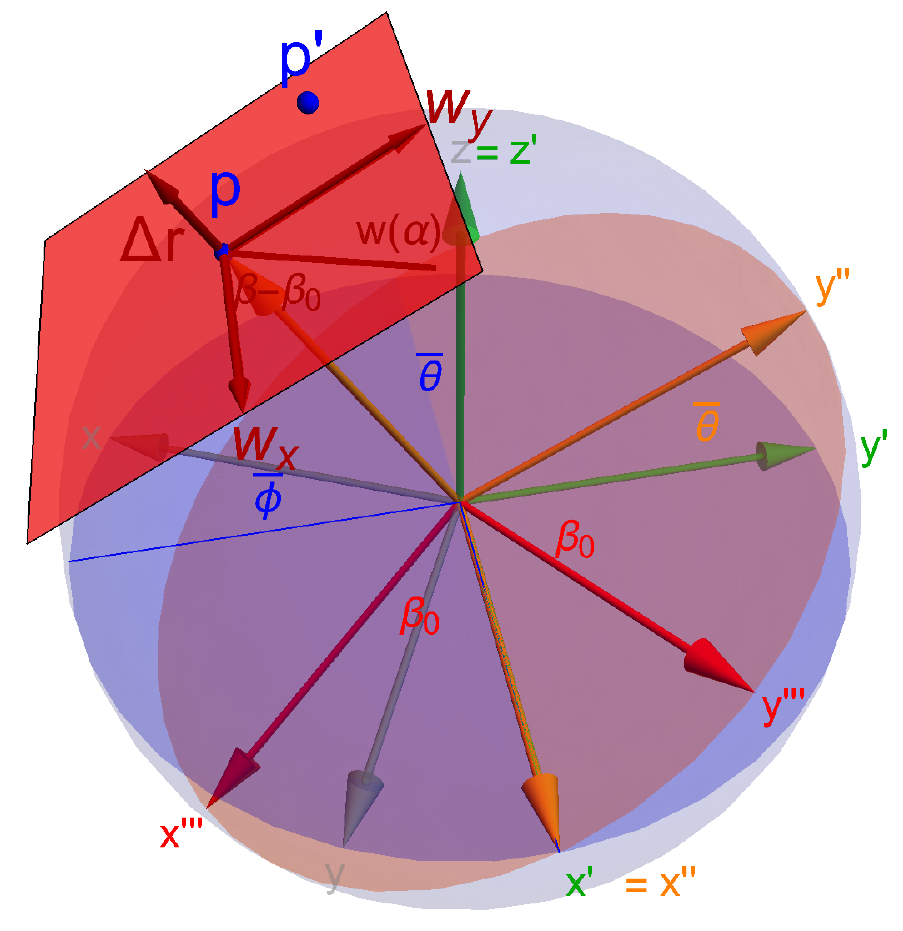}
\caption{Locally Cartesian coordinates $(w_x, w_y)$ are chosen to avoid ill behaviour on the pole. The particle is centred at $(0,0,0)$ in these coordinates and other points have have coordinates $(w_x,w_y,\Delta r)$. The axes of $(w_x, w_y)$  align with our second rotation axes $(x'', y'')$. }
\label{fig:rotw}
\end{figure}

We start with an Euler rotation with angles $(\phb+\frac{\pi}{2}, \thb, -\beta_0)$ where the first two angles are used to place the particle on the pole and kill the $m$-sum,
\begin{eqnarray} \label{eqn:coordRotn}
\sin{\alpha} \cos{(\beta-\beta_0)}&=&\sin{\theta} \sin{(\phi-\phb)},\\
\sin{\alpha} \sin{(\beta-\beta_0)}&=&\cos{\theta} \sin{\thb}-\sin{\theta}  \cos{\thb} \cos{(\phi-\phb)},\nonumber \\
\cos{\alpha}&=&\cos{\theta} \cos{\thb}+\sin{\theta} \sin{\thb} \cos{(\phi-\phb)}. \nonumber
\end{eqnarray}
As these are not well-behaved on the pole (where we place our particle), as has become standard from \cite{Barack:2002mha},we transform further to `locally Cartesian coorinates',
\begin{eqnarray} 
w_x&=w(\alpha) \cos(\beta-\beta_0)&=2 \sin \left(\frac{\alpha}{2}\right) \cos(\beta-\beta_0), \label{eqn:wx} \\
w_y&=w(\alpha) \sin(\beta-\beta_0)&=2 \sin \left(\frac{\alpha}{2}\right) \sin(\beta-\beta_0),\label{eqn:wy}
\end{eqnarray}
as illustrate in fig.~\ref{fig:rotw}

The third angle $\beta_0$, with
\begin{equation}\label{eqn:cossin}
\sin 2\beta_0 = -\frac{2\mu}{\eta}, \quad \cos 2 \beta_0 = \frac{ \zeta^2 - \xi^2}{\eta}, \\
\end{equation}
where
\begin{eqnarray}
\zeta^2&\equiv&g_{w_x w_x}+u_{w_x w_x}, \\
\xi^2&\equiv&g_{w_y w_y}+u_{w_y w_y}, \nonumber \\
\mu&\equiv&g_{w_x w_y}+u_{w_x w_y} , \nonumber \\
\eta^2&=&4\mu^2+(\xi^2 - \zeta^2)^2,
\end{eqnarray}
is chosen to ensure
\begin{equation} \label{eqn:rho00}
\rho(0,0,w_x,w_y)^2 = 4 \sin^2 \left(\frac{\alpha}{2}\right) \left[
	X+Y \sin^2 \beta
	\right],
\end{equation}
where $X$ and $Y$ are simple functions of $\xb$. We consider $\rho$ with $\Delta r \rightarrow 0$ as it was previously shown in Appendix  B of \cite{Barack:2002mha}, that one switch the order of the limit and integration in Eq.~\eqref{eqn:FRS} for all orders after the leading term. The above rotation then produces
\begin{eqnarray}
\rho(w_x,w_y)^2 &=& 2 \sin^2  \frac{\alpha}{2} \left(\xi^2 + \zeta^2+\eta \right)\left[
	1-k(\xb) \sin^2 \beta
\right], \nonumber \\
&=& 4 \sin^2  \frac{\alpha}{2} \left(\frac{\eta}k \right)\chi,
\end{eqnarray}
where
\begin{eqnarray} \label{eqn:k}
\chi  (\xb, \beta)&=& 1-k(\xb) \sin^2 \beta, \\
k(\xb) &\equiv& \frac{2 \eta}{\xi^2 + \zeta^2 +\eta}.
\end{eqnarray}
$F^{\rm{\sing}}_a$ of Eq.~\eqref{eqn:fasum} is now be rewritten as
\begin{widetext}
\begin{eqnarray} \label{eqn:fasum00}
F^{\rm{\sing}}_a \left(0,0,\alpha,\beta\right) &=&
 \sum_{n=-1}^{\infty} \sum_{p=-n-2}^{\lfloor (n-1)/2\rfloor} \sum_{q=0}^{n-2p} b^{[n]}_{\ab (n-2p,q)}(\bar{x}) \left(2\sin^2 \frac{\alpha}{2}\right)^{(n-1)/2} 2^{(n-1)/2} \binom{n-2p}{q}
\nonumber \\ && \qquad \qquad \qquad \times
\cos^{q}(\beta-\beta_0) \sin^{(n-2p-q)} (\beta-\beta_0) 
	\left(\frac{\eta \chi }k\right)^{(2 p - 1)/2} \epsilon^{n-1}, 
\end{eqnarray}
\end{widetext}
with the same caveat of the limits of the $p$-sum changing for $n=2$. Here as in Paper I the  $b^{[n]}_{\ab (n-2p,q)}$ coefficient has been relabelled
\begin{equation}
b^{[n]}_{\ab (n-2p, q)} \equiv b^{[n]}_{\ab \bar{c}_1 \dots \bar{c}_{(n-2p)}},
\end{equation}
where $n-2p$ is the total number of $w_x$'s and $w_y$'s in the sequence $\bar{c}_1 \dots \bar{c}_{(n-2p)}$ and $q$ the number of $w_x$'s.

To integrate over $\alpha$, one expands $ \left(2\sin^2 \frac{\alpha}{2}\right)^{(n-1)/2} $ in Legendre polynomials and makes use of their orthogonality, as well as several key formulas outline in Appendix D of \cite{Detweiler:2002gi} and reviewed in Paper I. The $\beta$ integral uses the definitions of $\chi$, Eq.~\eqref{eqn:chi} to capture all $\beta$ dependence in $\chi$ with
\begin{eqnarray}
\cos^2 (\beta)&=&\frac1k\left(
	k-1+\chi \right) , \\
\sin^2 (\beta)&=&\frac1k\left(
	1-\chi
\right). \nonumber
\end{eqnarray}
As was done in  \cite{Detweiler:2002gi}, we integrate the powers of $\chi$ using
\begin{eqnarray} \label{eqn:zeta}
\frac{1}{2 \pi} \int  \frac{d \beta}{\chi(\beta)^{n/2}} &=& 
\left<\chi^{-{n/2}}(\beta) \right> 
={}_2 F_1 \left(\frac{n}{2}, \frac{1}{2}; 1; k \right), \nonumber
\end{eqnarray}
where $(n+1)/2\in\mathbb{N}\cup \{0\}$ and ${}_2 F_1$ are hypergeometric functions. The recurrence
relation in Eq.~(15.2.10) of \cite{Abramowitz:Stegun},
\begin{equation}
\mathcal{F}_{p+1} (k) = \frac{p-1}{p \left(k - 1\right)} \mathcal{F}_{p-1}(k) + \frac{1 - 2p + \left(p - \frac{1}{2}\right) k}{p \left(k - 1\right)} \mathcal{F}_p(k),
\end{equation}
is then used to reduce the number of hypergeometric functions to two, ${}_2F_1 \left(\pm\frac{1}{2}, \frac{1}{2};1;k \right)$. These in turn translate to elliptic integrals
\begin{IEEEeqnarray}{lClClCl}
\left<\chi^{-\frac{1}{2}}\right> 
&=& {}_2F_1 \left(\frac{1}{2}, \frac{1}{2};1;k \right) &=& \frac{2}{\pi} \mathcal{K}, \\
\left<\chi^{\frac{1}{2}}\right> 
&=& {}_2F_1 \left(-\frac{1}{2}, \frac{1}{2};1;k \right) &=& \frac{2}{\pi} \mathcal{E} ,
\end{IEEEeqnarray}
where
\begin{eqnarray}
\mathcal{K} &\equiv& \int_0^{\pi/2} (1 - k \sin^2 \beta)^{-1/2} d\beta,  \\
\mathcal{E} &\equiv& \int_0^{\pi/2} (1 - k \sin^2 \beta)^{1/2} d\beta,
\end{eqnarray}
are complete elliptic integrals of the first and second kinds, respectively. 

\subsection{Results}
As was noted in Paper I, the techniques outlined here will not produce the leading order parameter previously produced by Barack and Ori \cite{Barack:2002mh}. Our rotation was designed to kill the $w_x w_y$ cross term in $\rho$, Eq.~\eqref{eqn:rhoS}. For the leading term, one cannot set $\Delta r \rightarrow 0$ and therefore has other problematic cross terms $\Delta r w_x$ and $\Delta r w_y$. The integration techniques described here will thus fail. Barack and Ori dealt with these terms by targetting $u_{w_y} \rightarrow 0$ as this is the source of two of these terms, the remaining $\Delta r w_x$ term is integrated by scaling the integrals with $\Delta r$ and expanding out the Legendre polynomial. These don't easily extend to higher orders, which is why we chose a different rotation and adapted the integration of techniques of Detweiler et al. \cite{Detweiler:2002gi}.

We reiterate the leading parameter here as defined by Barack and Ori \cite{Barack:2002mh}, as there was a typo in the parameters in the following review \cite{Barack:2009}
\begin{eqnarray}
F_{t[-1]} &=& -\frac{u^{\rb}}{u^{\tb}}F_{r[-1]}, \quad F_{\theta [-1]} =0, \quad F_{\phi [-1]} = 0, \nonumber \\
F_{r[-1]} &=&\mp \frac{\sgn \Delta r}{2 V}
\sqrt{\frac{\sin ^2\thb }{g_{\phb \phb} \Delta}}
   \sqrt{V +\frac{\Delta  u_{\rb}^2}{\Sigma}}, \nonumber 
\end{eqnarray}
where $\mp$ of $F_{r[-1]} $ is $-1$ for electromagnetism and a $+1$ for gravity,
\begin{equation}
V = 1 + \frac{u_{\thb}^2}\Sigma + \frac{L^2}{g_{\phb \phb}}, \nonumber
\end{equation}
with all having the same $\ell$ dependency is $\mathcal{A}^\ell_{[-1]}=1+2 \ell$.

The higher parameters produced for this work are too large to print, we therefore make them publicly available via a Mathematica package on Zenodo \cite{heffernan_anna_2022_7071504}. As the only generic Kerr gravity calculation is not in the Lorenz gauge, there is no data currently available to test whether the parameters give the expected convergence (as was shown in Paper I for the scalar case). However, as shown in the readme notebook accompanying the package, we do show that we successfully retrieve the parameters of Schwarzschild \cite{Heffernan:2012su} and equatorial Kerr {Heffernan:2012vj} which were shown to successfully regularise.

\section{Discussion} \label{sec:dis}
In this work, we produced the first high order (usually referring to anything after the first two) regularisation parameters for generic orbits in Kerr spacetime for both the electromagnetic and gravitational cases. As LISA approaches, there is currently one code that has successfully implemented generic orbits in Kerr for gravity \cite{vandeMeent:2017bcc}. More and faster codes are needed and we hope this work will assist in such developments. The effect of this one additional parameter in the scalar case saw an improvement from a week to a day in code time \cite{Nasipak:2022xjh} and allows one to capture resonances to the desired accuracy.

Future work has many directions, one can build on previous non-geodesic work \cite{Heffernan:2017cad}, implementing in Kerr. There is also work on other gauges that has a lot of potential \cite{Thompson:2018lgb}. As well as looking at any increase efficiency that might be gained from using spin-weighted  \cite{Warburton:2013lea}.or spheroidal harmonics.

\section*{Acknowledgements}
The author would like to thank Maarten Van de Meent for sharing his calculations to ensure these results agreed with those previous. This work was greatly assisted by the open software Mathematica packages xTensor and xCoba \cite{Martin-Garcia:2007bqa, Martin-Garcia:2008yei} and hence the author would like to thank all those who give their time to its development. The author gratefully acknowledges funding from the European Union's Horizon 2020 research and innovation programme under grant agreement 661705-GravityWaveWindow. This work was supported by European Union FEDER funds, the Spanish Ministerio de Ciencia e Innovación, and the Spanish Agencia Estatal de Investigación grant PID2019-106416GB-I00/AEI/MCIN/10.13039/501100011033, as well as Comunitat Autònoma de les Illes Balears through the Conselleria de Fons Europeus, Universitat i Cultura and the
Direcció General de Política Universitaria i Recerca with funds from the Tourist Stay Tax Law ITS 2017-006
(PRD2018/24, PDR2020/11).


\bibliographystyle{apsrev4-1}
\bibliography{references}

\end{document}